%% file: main.tex
\pdfoutput=1
\documentclass{llncs}
\input{prelude}
\begin{document} 
\title{Relational reasoning via probabilistic coupling}
% \titlerunning{}  % abbreviated title (for running head)

\author{Gilles Barthe\inst{1} \and Thomas Espitau\inst{1,2} \and 
Benjamin Grégoire\inst{3} \and Justin Hsu\inst{4} \and 
\\Léo Stefanesco\inst{1,5} \and Pierre-Yves Strub\inst{1}}
%
% \authorrunning{} % abbreviated author list (for running head)
% \tocauthor{}
%
\institute{
  $\mbox{}^1$ IMDEA Software \qquad
  $\mbox{}^2$ ENS Cachan \qquad
  $\mbox{}^3$ Inria \\
  $\mbox{}^4$ University of Pennsylvania \qquad
  $\mbox{}^5$ ENS Lyon
}

\maketitle

% Modify the bibliography environment to call for the author-year
% system. This is done normally with the citeauthoryear option
% for a particular contribution.
\makeatletter
%% \renewenvironment{thebibliography}[1]
%%      {\section*{\refname}
%%       \small
%%       \list{}%
%%            {\settowidth\labelwidth{}%
%%             \leftmargin\parindent
%%             \itemindent=-\parindent
%%             \labelsep=\z@
%%             \if@openbib
%%               \advance\leftmargin\bibindent
%%               \itemindent -\bibindent
%%               \listparindent \itemindent
%%               \parsep \z@
%%             \fi
%%             \usecounter{enumiv}%
%%             \let\p@enumiv\@empty
%%             \renewcommand\theenumiv{}}%
%%       \if@openbib
%%         \renewcommand\newblock{\par}%
%%       \else
%%         \renewcommand\newblock{\hskip .11em \@plus.33em \@minus.07em}%
%%       \fi
%%       \sloppy\clubpenalty4000\widowpenalty4000%
%%       \sfcode`\.=\@m}
%%      {\def\@noitemerr
%%        {\@latex@warning{Empty `thebibliography' environment}}%
%%       \endlist}
%%       \def\@cite#1{#1}%
%%       \def\@lbibitem[#1]#2{\item[]\if@filesw
%%         {\def\protect##1{\string ##1\space}\immediate
%%       \write\@auxout{\string\bibcite{#2}{#1}}}\fi\ignorespaces}

\makeatother

\begin{abstract}
  Probabilistic coupling is a powerful tool for analyzing pairs of probabilistic
  processes. Roughly, coupling two processes requires finding an appropriate
  witness process that models both processes in the same probability space.
  Couplings are powerful tools proving properties about the relation between two
  processes, include reasoning about convergence of distributions and
  \emph{stochastic dominance}---a probabilistic version of a monotonicity
  property.

  While the mathematical definition of coupling looks rather complex
  and cumbersome to manipulate, we show that the relational program
  logic pRHL---the logic underlying the EasyCrypt cryptographic proof
  assistant---already internalizes a generalization of probabilistic coupling.
  With this insight, constructing couplings is no harder than constructing
  logical proofs.  We demonstrate how to express and verify classic examples of
  couplings in pRHL, and we mechanically verify several couplings in EasyCrypt.
\end{abstract}

\section{Introduction}

\emph{Probabilistic couplings}~\citep{Thorisson00,Lindvall02,Villani08}
are a powerful mathematical tool for reasoning about pairs of
\emph{probabilistic processes}: streams of values that evolve randomly
according to some rule. While the two processes may be difficult to
analyze independently, a probabilistic coupling arranges processes $\{
u_i \}$, $\{ v_i \}$ in the same space---typically,
by viewing the pair of processes as randomly evolving
\emph{pairs} of values $\{ (u_i, v_i) \}$, coordinating the samples so that each
pair of values are related. In this way, couplings can reason about the relation
between the two processes.

From the point of view of program verification, a coupling is
a \emph{relational} program property, since it describes the relation between
two programs (perhaps one program run on two different inputs, or two completely
different programs). However, couplings are particularly interesting
for several reasons.

\paragraph*{Useful consequences.}

Couplings imply many other relational properties, and are
a powerful tool in mathematical proofs.

A classic use of coupling is showing that the distribution of the
value of two random processes started in different locations
eventually converges to the same distribution if we run the processes
long enough. This property is a kind of \emph{memorylessness}---or
\emph{Markovian}---property: The long-term behavior of the process is 
independent of its starting point. To prove memorylessness, the
typical strategy is to
couple the two processes so that their values move closer together; 
once the values meet, the two processes move together, yielding 
the same distribution. 

A different use of couplings is showing that one (numeric-valued)
process is, in some sense, bigger than the other. This statement has
to be interpreted carefully---since both processes evolve
independently, we can't guarantee that one process is
always larger than the other on all traces. \emph{Stochastic domination}
turns out to be the right definition: for any $k$, we require
$\Pr{}{u \geq k} > \Pr{}{v \geq k}$. This property follows if we can
demonstrate a coupling of a particular form.

\paragraph*{Relational from non-relational.}

Often, the behavior of the second coupled process is completely
specified by the behavior of the first; for instance, the second process may
mirror the first process. In such cases, the coupling allows us to
reason just about the first process. In other words, a coupling allows
us to prove certain relational properties by proving properties of a single
program.

\paragraph*{Compositional proofs.}

Typically, couplings are proved by coordinating corresponding samples of the two
processes, step by step; paper proofs call this process ``building a coupling'',
reflecting the piecewise construction of the coupled distribution. As a result,
couplings
can be proved locally by considering small pieces of the programs in isolation,
enabling
convenient mechanical verification of couplings.

\subsection*{Contributions}
In this paper, we apply
relational program verification to probabilistic couplings. While the
mathematical definition of coupling is seemingly
far from program verification technology, our primary insight is that
the logic pRHL from~\citet{BartheGZ09} already internalizes coupling in
disguise. More precisely, pRHL is built around a \emph{lifting}
construction, which turns a relation $R$ on two sets $A$ and
$B$ into a relation $R^\dagger$ over the set of sub-distributions over
$A$ and the set of sub-distributions over $B$.  Two programs are
related by $R^\dagger$ precisely when there exists a coupling of their
output sub-distributions whose support only contains pairs of values
$(u, v)$ which satisfy $R$.

This observation has three immediate consequences. First, by selecting
the relation $R$ appropriately, we can express a wide variety of
coupling properties, like distribution equivalence and stochastic
domination. Second, by utilizing the proof system of pRHL, we can
constructing and manipulate couplings while abstracting away the mathematical
details.
Finally, we can leverage EasyCrypt, a proof assistant implementing
pRHL, to mechanically verify couplings.

\section{Preliminaries}

\paragraph*{Probabilistic coupling.}
We begin by giving an overview of probabilistic coupling. As we described
before, a coupling places two probabilistic processes (viewed as probability
distributions) in the same probabilistic space.

We will work with sub-distributions over discrete (finite or
countable) sets. A sub-distribution $\mu$ over a discrete set $A$ is
a function $A\rightarrow [0,1]$ such that $\sum_{a\in A}\mu(a)
\leq 1$, and its support $\mathsf{supp}(\mu)$ is the pre-image of
$(0,1]$. We let $\Dist(A)$ denote the set of sub-distributions over
  $A$. Every sub-distribution can be given a monadic structure;
  the unit operator maps every element $a$ in
  the underlying set to its Dirac distribution $\delta_a$ and the
  monadic composition $\mathsf{Mlet}(\mu,F) \in\Dist(B)$ of
  $\mu\in\Dist(A)$ and $F:A\rightarrow \Dist(B)$ is 
  $\mathsf{Mlet}(\mu,F)(b)=\sum_{a\in A} \mu(a) \times F(a)(b)$.

When working with sub-distributions over tuples, the probabilistic versions of
the usual projections on tuples are called \emph{marginals}. The first and
second marginals $\pi_1(\mu)$ and $\pi_2(\mu)$ of a distribution $\mu$ over
$A\times B$ are defined by $\pi_1 (\mu)(a)=\sum_{b\in B} \mu(a,b)$ and $\pi_2
(\mu)(b)=\sum_{a\in A} \mu(a,b)$. 
We can now formally define coupling.
\begin{definition}
The \emph{Frechet class} $\mathfrak{F}(\mu_1,\mu_2)$ of two
sub-distributions $\mu_1$ and $\mu_2$ over $A$ and $B$ respectively is
the set of sub-distributions $\mu$ over $A\times B$ such that
$\pi_1(\mu)=\mu_1$ and $\pi_2(\mu)=\mu_2$. Two sub-distributions $\mu_1,
\mu_2$ are said to be \emph{coupled} with witness $\mu$ if
$\mu\in\mathfrak{F} (\mu_1,\mu_2)$, i.e.\, $\mu$ is in the Frechet
class of $\mu_1, \mu_2$.
\end{definition}

\paragraph*{Lifting relations.}
Before introducing pRHL, we describe the \emph{lifting}
construction. This operation allows pRHL to make statements about
pairs of (sub-)distributions, and is a generalized form of
probabilistic coupling.

The idea is to define a family of couplings based on the support of
the witness distribution.  Given a relation $R\subseteq A\times B$ and
two distributions $\mu_1$ and $\mu_2$ over $A$ and $B$ respectively,
we let $\mathfrak{L}_R(\mu_1,\mu_2)$ denote the subset of
sub-distributions $\mu\in\mathfrak{F}(\mu_1,\mu_2)$ such that
$\mathsf{supp}(\mu) \subseteq R$.  Given a ground relation $R$, we view
distributions in $\mathfrak{L}_R$ as witnesses for a \emph{lifted}
relation on distributions.

\begin{definition}
The \emph{lifting} of a relation $R\subseteq A\times B$ is the relation
$R^\dagger\subseteq \Dist(A)\times\Dist(B)$ with
$\mu_1~R^\dagger~\mu_2$ iff $\mathfrak{L}_R(\mu_1,\mu_2)\neq \emptyset$.
\end{definition}
%% Lifting can also be characterized inductively, as the smallest
%% relation that contains the Dirac distributions of elements related by
%% $R$, and closed under convex combinations; that is, it is equivalent
%% to define $R^\dagger$ by the clauses:
%% %
%% \[
%%   \begin{array}{c@{~~~~~~~~~~~~~}c}
%%     \inferrule{a~R~b}{\delta_a~R^\dagger \delta_b} &
%%     \inferrule{\mu_i~R~\mu'_i \\ \sum_{i\in I} p_i=1}{
%%       \left(\sum_{i\in I} p_i~\mu_i\right) ~R^\dagger ~
%%     \left(\sum_{i\in I} p_i~\mu_i'\right)}  .
%%   \end{array}
%% \]
%% %

Before turning to the definition of pRHL, we give some intuition for
why lifting is useful. Roughly, if we know two distributions are
related by a lifted relation $R^\dagger$, we can treat two samples
from the distribution as if they were related by $R$. In other words,
the lifting machinery gives a powerful way to translate between
information about distributions and information about
samples. \citet{DengD11} provide an excellent introductory
exposition to lifting, and give several equivalent characterizations
of lifting.

\subsection{A pRHL primer}
We are now ready to present pRHL, a relational program logic for
probabilistic computations. In its original form~\citep{BartheGZ09},
implemented in the EasyCrypt proof
assistant~\citep{BartheGHZ11}, pRHL reasons about programs written in an
imperative language extended with random assignments with the following
syntax of commands:
\[
  c ::= \Ass{x}{e} \mid \Rand{x}{d} \mid \Cond{e}{c}{c} \mid \WhileL{e}{c} \mid
  \Skip \mid \Seq{c}{c}
\]
where $e$ ranges over expressions, $d$ ranges over distribution
expressions, and $\Rand{x}d$ stores a sample from $d$ into $x$. Commands are
interpreted as functions from memories to
distributions over memories; using the fixed point theorem for Banach
spaces, one can define for each command $c$ a function $\Sem{c}:\Mem
\rightarrow \Dist(\Mem)$, where $\Mem$ is the set of well-typed maps
from program variables to values.

Assertions in the language are first-order formulae over generalized
expressions. The latter are built from tagged variables $x_1$ and
$x_2$, which correspond to the interpretation of the
program variable $x$ in the first and second memories.
Assertions in pRHL are deterministic and do not refer
to probabilities.
\begin{definition}
A \emph{pRHL judgment} is a quadruple of the form
$\Equiv{c_1}{c_2}{\pre}{\post}$, where $\pre$ and $\post$ are
assertions, and $c_1$ and $c_2$ are separable statements, i.e.\, they
do not have any variable in common.
A judgment is \emph{valid} iff for all memories $m_1$ and $m_2$, we have
$(m_1,m_2)\models\pre \Rightarrow
(\Sem{c_1}(m_1),\Sem{c_2}(m_2))\models\post^\dagger$ .
\end{definition}
Judgments can be proved valid with a variety of rules.
% \begin{itemize}
% \item \emph{Two-sided rules} operate on the two programs at
%   the same time, requiring that both programs have the same shape and follow the
%   same control flow (\Cref{fig:rules:twosided}). 
% \item \emph{One-sided rules} operate on one program at the time
%   (\Cref{fig:rules:onesided}).
% \item \emph{Structural rules} apply independent of the shape of the program
%   (\Cref{fig:rules:struct}).
% \item \emph{Program transformation} rules allow replacing
%   one of the two programs in the judgment by a semantically equivalent
%   one (\Cref{fig:rules:struct}).
% \end{itemize}

\paragraph*{Two-sided and one-sided rules.}
The pRHL logic features two-sided rules (\Cref{fig:rules:twosided})
and one-sided rules (\Cref{fig:rules:onesided}). Roughly speaking,
two-sided rules relate two commands with the same structure and control flow, while
one-sided rules relate two commands with possibly different structure or control
flow; the latter rules allow pRHL to express \emph{asynchronous} couplings
between programs that may exhibit different control flow.

We point out two rules that will be especially important for our
purposes. The rule \rname{Sample} is used for relating two sampling
commands. Note that it requires an injective function $f : T_1 \to
T_2$ from the domain of the first sampling command to the domain of
the second sampling command. When the two sampling commands have the
same domain---as will be the case in our examples---$f$ is simply a
bijection on $T = T_1 = T_2$. This bijection gives us the freedom to
specify the relation between the two samples when we couple the samples.

The rule \rname{While} is the standard while rule adapted to
pRHL. Note that we require the guard of the two commands to be
equal---so in particular the two loops must make the same number of
iterations---and $\post$ plays the role of the while loop invariant as
usual.

\begin{figure}[t]
$$
\begin{array}{c}
% \inferrule*[Left=\textrm{Skip}]{~}{\Equiv{\Skip}{\Skip}{\post}{\post}}
% \\[2ex]

% \inferrule*[Left=\textrm{Seq}]{\Equiv{c_1}{c_2}{\post}{\post'} \\
%          \Equiv{c_1'}{c_2'}{\post'}{\post''}}
%         {\Equiv{c_1;c_1'}{c_2;c_2'}{\post}{\post''}}
% \\[2ex]

% \inferrule*[Left=\textrm{Ass}]{~}{\Equiv{\Ass{x_1}{e_1}}{\Ass{x_2}{e_2}}
%      {\post~[e_1/x_1,e_2/x_2]} {\post}}
% \\[2ex]
\inferrule*[Left=\textrm{Sample}]{f\in T_1\stackrel{1-1}{\longrightarrow} T_2 \\
\forall v\in T_1. ~d_1(v)=d_2(f~v)
}{
 \Equiv{\Rand{x_1}{d_1}}{\Rand{x_2}{d_2}}
       {\forall v, \post [v/x_1,f(v)/x_2]}
       {\post}}
\\[2ex]
\inferrule*[Left=\textrm{If}]{
\pre \Rightarrow e_1 = e_2
\\
\Equiv{c_1}{c_2}{\pre \land e_1}{\post}
\\
\Equiv{c_1'}{c_2'}{\pre \land \neg e_1}{\post}
}
{
\Equiv{\Cond{e_1}{c_1}{c_1'}}{\Cond{e_2}{c_2}{c_2'}}{\pre}{\post}
}
\\[2ex]
\inferrule*[Left=\textrm{While}]{
\post \Rightarrow e_1 = e_2
\\
\Equiv{c_1}{c_2}{\post \land e_1}{\post}
}
{
\Equiv{\WhileL{e_1}{c_1}}{\WhileL{e_2}{c_2}}
 {\post}
 {\post \land \neg e_1}
}
\end{array}
$$
\caption{Two-sided proof rules (selection)}
\label{fig:rules:twosided}
\end{figure}

\begin{figure}[t]
$$\begin{array}{c}
% \inferrule*[Left=\textrm{AssL}]{
% \Equiv{\Skip}{c}{\pre [e_1/x_1]}{\post}
% }{
% \Equiv{\Ass{x_1}{e_1}}{c}{\pre}{\post}
% } 
% \\[2pt]
\inferrule*[Left=\textrm{SampleL}]{\Equiv{\Skip}{c}{\forall v,\pre[v/x_1]}{\post}}{
 \Equiv{\Rand{x_1}{d_1}}{c}{\pre}{\post}}
\\[2pt]
\inferrule*[Left=\textrm{IfL}]{ 
\Equiv{c_1}{c}{\pre\land e_1}{\post} 
  \and 
  \Equiv{c_1'}{c}{\pre \land\lnot e_1}{\post}}{
  \Equiv{\Cond{e_1}{c_1}{c_1'}}{c}{\pre}{\post}
}
\\[2pt]
\inferrule*[Left=\textrm{WhileL}]{ 
\Equiv{c_1}{\Skip}{\post\land e_1}{\post} 
  \and  \WhileL{e_1}{c_1}~\mathrm{lossless}}{
  \Equiv{\WhileL{e_1}{c_1}}{\Skip}{\post}{\post}
}
\end{array}
$$
\caption{One-sided proof rules (selection)}
\label{fig:rules:onesided}
\end{figure}

\paragraph*{Structural and program transformation rules.}
pRHL also features structural rules that are very similar to those of
Hoare logic, including the rule of consequence and the case rule. In
addition, it features a rule for program transformations,
based on an equivalence relation $\simeq$ that provides a sound
approximation of semantical equivalence. For our examples, it is
sufficient that the relation $\simeq$ models loop range splitting
and biased coin splitting, as given by the following clauses:
$$\begin{array}{rcl}
\WhileL{e}{c} & \simeq & \WhileL{e\wedge e'}{c};\WhileL{e}{c} \\
\Rand{x}{\bernD(p_1 \cdot p_2)} & \simeq & 
\Rand{x_1}{\bernD(p_1)};\Rand{x_2}{\bernD(p_2)};\Ass{x}{x_1\wedge x_2}
\end{array}
$$
\Cref{fig:rules:struct} provides a selection of structural and
program transformation rules.
\begin{figure}
$$\begin{array}{c}
\inferrule*[Left=\textrm{Conseq}]{\Equiv{c_1}{c_2}{\pre'}{\post'}
 \\ \pre \Rightarrow \pre'
 \\ \post' \Rightarrow \post}
{\Equiv{c_1}{c_2}{\pre}{\post}}
\\[2ex]
\inferrule*[Left=\textrm{Case}]{
  \Equiv{c_1}{c_2}{\pre \land \pre'}{\post}
  \quad
  \Equiv{c_1}{c_2}{\pre \land \neg~\pre'}{\post}
}
{\Equiv{c_1}{c_2}{\pre}{\post}}
\\[2ex]
\inferrule*[Left=\textrm{Equiv}]{
  \Equiv{c_1'}{c_2'}{\pre}{\post} \\ c_1\simeq c_1' \\ c_2 \simeq c_2'}
{\Equiv{c_1}{c_2}{\pre}{\post}}
\end{array}$$
\caption{Structural and program transformation rules (selection)}
\label{fig:rules:struct}
\end{figure}

\subsection{From pRHL judgments to probability judgments}
We will derive two kinds of program properties
from the existence of an appropriate probabilistic coupling. We will
first discuss the mathematical theorems, where the notation is lighter
and the core idea more apparent, and then demonstrate how the
mathematical version can be expressed in terms of pRHL judgments.

\paragraph*{Total variation and coupling.}
The first principle bounds the distance between two distributions in
terms of a probabilistic coupling. We first define the total variation
distance, also known as statistical distance, on distributions.

\begin{definition}
  Let $X$ and $X'$ be distributions over a countable set $A$. The
  \emph{total variation (TV) distance} between $X$ and $X'$ is defined
  by \\
  $\| X - X' \|_{tv} \triangleq \frac{1}{2} \sum_{a \in A} | X(a) - X'(a) |$ .
\end{definition}
% The normalization of $1/2$ is purely for notational convenience; it
% ensures that the TV distance between any two distributions is at most
% $1$.

To bound the distance between two distributions, it is enough to find
a coupling and bound the probability that the two coupled variables
differ. 
\begin{theorem}[Total variation, see~\cite{Lindvall02}]
  \label{thm:tv}
  Let $X$ and $X'$ be distributions over a countable set. Then for any
  coupling $Y = (\hat{X}, \hat{X'})$, we have
  \[
    \| X - X' \|_{tv} \leq \Pr {(x, x') \sim Y} { x \neq x' } .
  \]
\end{theorem}
This theorem is useful for reasoning about convergence of
distributions.

To describe a pRHL analog of this theorem, we first introduce some
useful notation. For all memories $m$ and expressions $e$, we write $m(e)$ for
the interpretation of $e$ in memory $m$. For all expressions $e$ of type $T$ and distribution
$\mu$ over memories, let $\Sem{e}_\mu$ be defined as $\textsf{Mlet}
~m=\mu ~\textsf{in}~\textsf{unit}~m(e)$; note that $\Sem{e}_\mu$
denotes a \emph{distribution} over $T$. Similarly, for all events $E$
(modeled as a boolean expression encoding a predicate over memories) and distribution $\mu$
over memories, let $\Sem{E}_\mu$ be defined as $\textsf{Mlet}~
m=\mu~\textsf{in}~\textsf{unit}~E(m)$. Thus, $\Sem{E}_\mu$ is
the probability of event $E$ holding in the distribution $\mu$. Then,
\Cref{thm:tv} can be written in terms of pRHL.
\begin{proposition}
If $\Equiv{c_1}{c_2}{\pre}{\post \Rightarrow v_1 = v_2}$, where
$\post$ exclusively refers to variables in $c_1$, then for all
initial memories $m_1$ and $m_2$ that satisfy the precondition, the
total variation distance between $\Sem{v_1}_{\Sem{c}(m_1)}$ and
$\Sem{v_2}_{\Sem{c}(m_2)}$ is at most
$\Sem{\neg\post}_{\Sem{c}(m_1)}$, i.e.\,
$ \| \Sem{v_1}_{\Sem{c_1}(m_1)} - \Sem{v_2}_{\Sem{c_2}(m_2)} \|_{tv} \leq 
  \Sem{\neg\post}_{\Sem{c}(m_1)}$ .
\end{proposition}
This proposition underlies the ``up-to-bad'' reasoning in EasyCrypt.

\paragraph*{Stochastic domination and coupling.}
A second relational property of distributions is \emph{stochastic domination}.
\begin{definition}
  Let $X$ and $X'$ be distributions over set $A$ with an order relation $\geq$.
  We say $X$ \emph{stochastically dominates} $X'$, written $X \geq_{sd} X'$, if
  for all $a \in A$,
  \[
    \Pr {x \sim X} {  x \geq a } \geq \Pr {x' \sim X'}{ x' \geq a } .
  \]
\end{definition}
Intuitively, stochastic domination defines a partial order on distributions over
$A$ given an order over $A$. Strassen's theorem shows that stochastic dominance
is intimately related to coupling.

\begin{theorem}[Strassen's theorem, see~\citet{Lindvall02}]
  Let $X$ and $X'$ be distributions over a countable ordered set $A$. Then $X
  \geq_{sd} X'$ if and only if there is a coupling $Y = (\hat{X}, \hat{X'})$
  with $Y \in \mathfrak{L}_{\geq}(X, X')$.
\end{theorem}

The forward direction is usually the more useful direction; we can express it in
the following pRHL form.
\begin{proposition}
  \label{thm:sd}
If $\Equiv{c_1}{c_2}{\pre}{v_1 \geq v_2}$, then for all initial
memories $m_1$ and $m_2$ that satisfy the precondition,
$\Sem{v_1}_{\Sem{c}(m_1)} \geq_{sd} \Sem{v_2}_{\Sem{c}(m_2)}$.
\end{proposition}

\section{Warming up: Random walks}

We warm up with couplings for random walks. These numeric processes
model the evolution of a token over a discrete space: at each time step 
the token will choose its next movement randomly. We will show that
if the two initial positions satisfy some property, the distributions of the two
positions converge.

\subsection{The basic random walk}

Our first example is a random walk on the integers. Starting at an initial
position, at each step we flip a fair coin. If heads, we move one step to the
right. Otherwise, we move one step to the left. The code for running process $k$
steps is presented in the left side of \Cref{fig:walk}.
The variable $\lstt{H}$ stores the history of coin flips. While this history
isn't needed for computation of the result (it is \emph{ghost code}), we will
state invariants in terms of this history.

\begin{figure}
  \begin{subfigure}[b]{0.5\textwidth}
\begin{lstlisting}
pos := start; H := []; i := 0;
while i < k do
  b ~~ {0,1};
  H := b :: H;
  if b then pos++ else pos-- fi;
  i := i + 1;
end
return pos

$\vphantom{}$
\end{lstlisting}
    \caption{Random walk on $\ZZ$}
  \end{subfigure}
  \vrule
  \begin{subfigure}[b]{0.5\textwidth}
\begin{lstlisting}
pos := start; H := []; i := 0;
while i < k do
  mov ~~ {0,1};
  dir ~~ {0,1};
  crd ~~ [1,d];
  H := (mov, dir, crd) :: H;
  if mov then
    pos := pos + (dir ? 1 : -1 ) * u(crd)
  fi;
  i   := i + 1;
end
return pos
\end{lstlisting}
    \caption{Random walk on $(\ZZ/k\ZZ)^d$}
  \end{subfigure}
  \caption{Two random walks}
  \label{fig:walk}
\end{figure}

We consider two walks that start at locations $\lstt{start}_1$ and
$\lstt{start}_2$ that are an even distance apart: $\lstt{start}_2 -
\lstt{start}_1 = 2n \geq 0$. We want to show that the distribution on end
positions in the two walks converges as $k$ increases. From \Cref{thm:tv}, it
suffices to find a coupling of the two walks, i.e., a way to coordinate their
random samplings.

The basic idea is to \emph{mirror} the two walks. When the first process moves
towards the second process, we have the second process also move closer; when
the first process moves away, we have the second process move away too. When the
two processes meet, we have the two processes make identical moves.

To carry out this plan, we define $\Sigma(\lstt{H})$ to be the number of
\textsf{true} in \lstt{H} minus the number of \textsf{false}; in terms of the
random walk, $\Sigma(\lstt{H})$ measures the net change in position of a process
with history $\lstt{H}$.  Then, we define a predicate such that $P(\lstt{H})$
holds when $\lstt{H}$ contains a prefix $\lstt{H'}$ such that $\Sigma(\lstt{H'})
= n$. 

Accordingly, $P(\lstt{H}_1)$ holds when the first process has moved at least $n$
spots to the right. Under the coupling, this means that the second process must
have moved at least $n$ spots to the left since the two particles are mirrored.
Since the first process starts out exactly $2n$ to the left of the second
process, $P(\lstt{H}_1)$ is true exactly when the coupled processes have already
met. If the processes start out an \emph{odd} distance apart, then they will
\emph{never} meet under this coupling---the coupling preserves the parity of the
distance between the two positions.

To formalize this coupling in pRHL, we aim to couple two copies of the program
above, which we denote $c_1$ and $c_2$. We relate the two \textsf{while} loops
with rule \rname{While} using the following invariant:
\[
  (\lstt{pos}_1 \neq \lstt{pos}_2
  \Rightarrow
  \lstt{pos}_1 = \lstt{i}_1 + \Sigma (\lstt{H}_1)
  \wedge
  \lstt{pos}_2 = \lstt{i}_2 - \Sigma (\lstt{H}_1))
  \wedge (P (\lstt{H}_1) \Rightarrow \lstt{pos}_1 = \lstt{pos}_2) .
\]
The loop invariant states that before the two particles meet, their
trajectories are mirrored, and that once they have met, they coincide forever.

To prove that this is an invariant, we need to relate the loop bodies. The key
step is relating the two sampling operations using the rule \rname{Sample}; note
that we must provide a bijection $f$ from booleans to booleans. We choose
the bijection based on whether the two coupled walks have met or not.

More precisely, we perform a case analysis on $\lstt{pos}_1 = \lstt{pos}_2$ with
rule \rname{Case}. If they are equal then the walks move together, so we use the
identity map for $f$; this has the effect of forcing both processes to see the
same sample.  If the walks are at different positions, we use the negation map
$(\neg)$ for $f$, so as to force the two processes to take opposite steps. 

Putting everything together, we can prove the following judgment in pRHL:
\[
  \Equiv{c_1}{c_2}{\lstt{start}_1 + 2n = \lstt{start}_2}{(P(\lstt{H}_1)
  \Rightarrow \lstt{pos}_1 = \lstt{pos}_2)} .
\]
By \Cref{thm:tv}, we can bound the TV distance between the final positions. If
two memories $m_1, m_2$ satisfy $m_1(\lstt{start}) + 2n = m_2(\lstt{start})$, we
have
\[
  \| \Sem{\lstt{pos}_1}_{\Sem{c_1}(m_1)} - \Sem{\lstt{pos}_2}_{\Sem{c_2}(m_2)}
  \|_{tv}
  \leq
  \Sem{\neg P(\lstt{H}_1)}_{\Sem{c_1}(m_1)} .
\]
Note that the right hand side depends only on the first program. In other
words, proving this quantitative bound on two programs is reduced to proving a
quantitative property on a \emph{single} program---this is the power of
coupling.

\subsection{Lazy random walk on a torus}

For a more interesting example of a random walk, we can consider a walk on a
torus. Concretely, the position is now a $d$-tuple of integers in $[0, k-1]$.
The walk first flips a fair coin; if heads it stays put, otherwise it moves. If
it moves, the walk chooses uniformly in $[1,d]$ to choose the coordinate to
move, and a second fair coin to determine the direction (positive, or negative).
The positions are cyclic: increasing from $k-1$ leads to $0$, and
decreasing from $0$ leads to $k-1$.

We can simulate this walk with the program in the right side of \Cref{fig:walk},
where $\lstt{u}(i)$ is the $i$-th canonical base vector in $(\ZZ/k\ZZ)^d$.  As
before, we store the trace of the random walk in the list $\lstt{H}$. All
arithmetic is done modulo $k$.

Like the simple random walk, we start this process at two locations
$\lstt{start}_1$ and $\lstt{start}_2$ on the torus and run for $k$ iterations.
We aim to prove that the distributions of the two walks converge as $k$
increases by coupling the two walks, iteration by iteration.
Each iteration, we first choose
the same coordinate \lstt{crd} and the same direction \lstt{dir} in both walks.
If the two positions coincide in coordinate \lstt{crd}, we arrange both walks to
select the same movement flag \lstt{mov}, so that the walks either move together, or both stay
put. If the two positions
differ in \lstt{crd}, we arrange the walks to select opposite samples
in \lstt{mov} so that exactly one walk moves.

As in the basic random walk, we can view our coupling as letting the first
process evolve as usual, then coordinating the samples of the second process to
perform the coupling. In other words, given a history $\lstt{H}_1$ of samples
for the first process, the behavior of the second coupled process is completely
specified.

Thus, we can define operators to extract the movements of each walk from the
trace $\lstt{H}_1$ of the samplings of the first process:
$\Sigma_1(i,\lstt{H}_1)$ is the drift of the $i$th coordinate of the
first process, and $\Sigma_2(i, \lstt{H}_1)$ is the drift of the second process.
Essentially, these operators encode the coupling by describing how the second
process moves as a function of the first process's samples.

In pRHL, we will use the rule \rname{While} with the following invariant:
\begin{align*}
  &\forall i \in [1, d].\; (\Sigma_1(i, \lstt{H}_1) - \Sigma_2(i, \lstt{H}_1)
  = \Delta[i] \Rightarrow \lstt{pos}_1[i] = \lstt{pos}_2[i]) \\
  & \mathrel{\wedge} (\lstt{pos}_1[i] \neq \lstt{pos}_2[i] \Rightarrow
  \lstt{pos}_1[i] = \lstt{start}_1[i] + \Sigma_1(i, \lstt{H}_1) \wedge
  \lstt{pos}_2[i] = \lstt{start}_2[i] + \Sigma_2(i, \lstt{H}_1)) ,
\end{align*}
where $\Delta$ is the vector $\lstt{start}_2 - \lstt{start}_1$.
The first conjunct states that the walks move together in coordinate $i$ once
they couple in coordinate $i$, while the second conjunct describes the
positions in terms of the history $\lstt{H}_1$.

To prove that the invariant is preserved, we encode the coupling described above
into pRHL, via three uses of the rule \rname{Sample}. The first two
samples---for \lstt{crd} and \lstt{dir}---are coupled with $f$ being
identity bijections (on $[1, d]$ and on booleans), ensuring that
the processes make identical choices. When
sampling \lstt{mov}, we inspect the history $\lstt{H}_1$ to see whether the two
walks agree in position \lstt{crd}. If so, we choose the identity bijection for
\lstt{mov}; if not, we choose negation. This coupling is sufficient to verify
the loop invariant.

To conclude our proof, the first conjunct in the invariant implies that we
can prove the pRHL judgment $\Equiv{c_1}{c_2}{\lstt{start}_2 - \lstt{start}_1 =
  \Delta}{\post}$, where
\[
  \post \triangleq (\forall i \in [1,
    d].\; \Sigma_1(i, \lstt{H}_1) - \Sigma_2(i, \lstt{H}_1)
  = \Delta[i]) \Rightarrow \forall i \in [1, d].\; \lstt{pos}_1[i] =
  \lstt{pos}_2[i] .
\]
Finally, \Cref{thm:tv} implies that for any two initial memories $m_1, m_2$ with
$m_2(\lstt{start}) - m_1(\lstt{start}) = \Delta$, we have
\[
  \| \Sem{\lstt{pos}_1}_{\Sem{c_1}(m_1)} - \Sem{\lstt{pos}_2}_{\Sem{c_2}(m_2)} \|_{tv}
  \leq
  \Sem{\exists i \in [1,
    d].\; \Sigma_1(i, \lstt{H}_1) - \Sigma_2(i, \lstt{H}_1)
    \neq \Delta[i]}_{\Sem{c_1}(m_1)} .
\]
Again, proving a quantitative bound on the convergence of two distributions is
reduced to proving a quantitative bound on a single program.

\section{Combining coupling with program transformation}

So far, we have seen examples where the coupling is proved directly on the two
original programs $c_1$ and $c_2$. Often, it is convenient to introduce
a third program $c^*$ that is equivalent to $c_1$, and then couple $c^*$ to
$c_2$. Applying transitivity (rule \rname{Equiv}), this gives a coupling
between $c_1$ and $c_2$. Let's consider two examples.

\subsection{Two biased coins}

Consider a coin flipping process that flips a coin $k$ times, and returns the
number of heads observed. We consider this process run on two different biased
coins: The first coin has probability $q_1$ of coming up heads, while the second
coin has probability $q_2$ of coming up heads with $q_1 \geq q_2$. Let the
distribution on the number of heads be $\mu_1$ and $\mu_2$ respectively.

Intuitively, it is clear that the first process is somehow bigger than the
second process: it is more likely to see more heads, since the first coin is
biased with a higher probability. Stochastic dominance turns out to be the
proper way to formalize our intuition.
To prove it, \Cref{thm:sd} implies that we just need to find an appropriate
coupling of the two processes.

\begin{figure}[t]
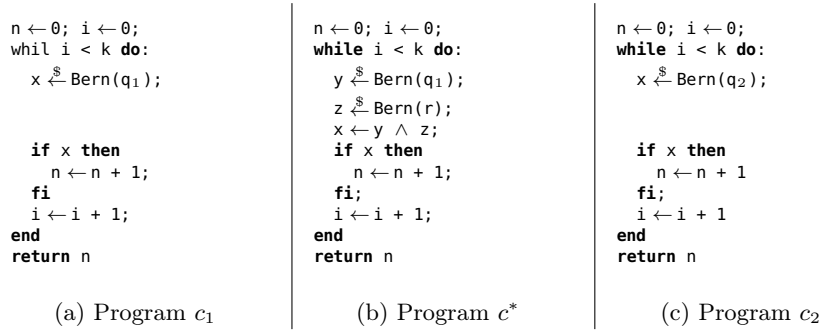

  \begin{subfigure}[t]{0.32\textwidth}
    \begin{lstlisting}
n := 0; i := 0;
whil i < k do:
  x ~~ Bern(q$_1$);
  $\vphantom{\lstrnd}$

  if x then
    n := n + 1;
  fi
  i := i + 1;
end
return n
    \end{lstlisting}
    \caption{Program $c_1$}
  \end{subfigure}
  \vrule
  \begin{subfigure}[t]{0.32\textwidth}
    \begin{lstlisting}
n := 0; i := 0;
while i < k do:
  y ~~ Bern(q$_1$);
  z ~~ Bern(r);
  x := y $\land$ z;
  if x then
    n := n + 1;
  fi;
  i := i + 1;
end
return n
    \end{lstlisting}
    \caption{Program $c^*$}
  \end{subfigure}
  \vrule
  \begin{subfigure}[t]{0.32\textwidth}
    \begin{lstlisting}
n := 0; i := 0;
while i < k do:
  x ~~ Bern(q$_2$);
  $\vphantom{\lstrnd}$

  if x then
    n := n + 1
  fi;
  i := i + 1
end
return n
    \end{lstlisting}
    \caption{Program $c_2$}
  \end{subfigure}
  \caption{Coupling for biased coin flips}
  \label{fig:biased-games}
\end{figure}

While it is possible to define a coupling
directly by carefully coordinating the corresponding coin flips, we will give a
simpler coupling that proceeds in two stages. First, we will couple a program
$c_1$ computing $\mu_1$ to an intermediate program $c^*$. Then, we will show that
$c^*$ is equivalent to a program $c_2$ computing $\mu_2$, thus exhibiting a
coupling between $\mu_1$ and $\mu_2$.  Letting $r = q_2/q_1$ and denoting the
coin flip distribution with probability $p$ of sampling true by $\bernD(p)$, we
give the programs in \Cref{fig:biased-games}. 

For the first step, we want to couple $c_1$ and $c^*$. For a rough sketch, we
want to use rule \rname{While} with an appropriate loop invariant; here, $n_1
\geq n^*$. To show that the invariant is preserved, we need to relate the loop
bodies.  We use the two-sided rule \rname{Sample} when sampling $\lstt{x}$ and
$\lstt{y}$ (taking the bijection $f$ to be the identity), the one-sided rule
\rname{Sample-L} to relate sampling nothing (\textsf{skip}) in $c_1$ with sampling $\lstt{z}$
in $c^*$, and the one-sided rule \rname{IfL} to relate the two conditionals. (The
one-sided rule is needed, since the two conditionals may take different
branches.) Thus, we can prove the judgment $\Equiv{c_1}{c^*}{q_1 \geq q_2 \land r =
  q_2/q_1}{n_1 \geq n^*}$.

For the second step, we need to prove that $c^*$ is equivalent to $c_2$. Here,
we use a sound approximation $\simeq$ to semantic equivalence as described
in the preliminaries. Specifically, we have
$ \Rand{\lstt{x}}{\bernD(q_1 \cdot r)}
  \simeq 
  \Rand{\lstt{y}}{\bernD(q_1)};\Rand{\lstt{r}}{\bernD(r)};\Ass{\lstt{x}}{\lstt{y}
    \wedge \lstt{z}} $
for the loop bodies; showing equivalence of $c^*$ and
$c_2$ is then straightforward. Thus, we can show $\Equiv{c^*}{c_2}{q_1 \geq q_2
  \land r = q_2/q_1}{n^* = n_2}$. Applying rule \rname{Equiv} gives the final
judgment $\Equiv{c_1}{c_2}{q_1 \geq q_2 \land r = q_2/q_1}{n_1 \geq n_2}$,
showing stochastic domination by \Cref{thm:sd}.

\subsection{Balls into bins: asynchronous coupling}

\begin{figure}
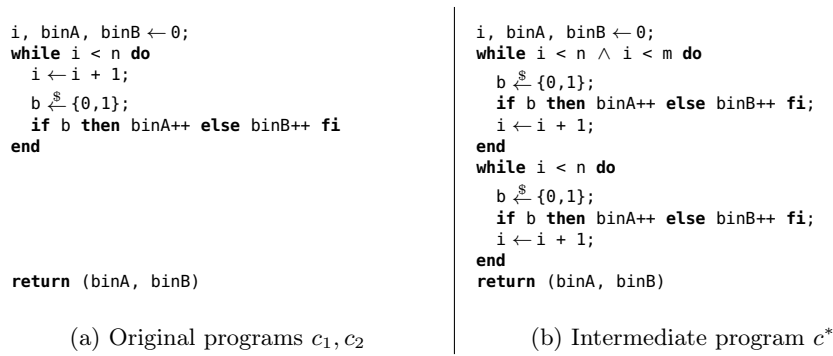

  \begin{subfigure}[t]{0.5\textwidth}
\begin{lstlisting}
i, binA, binB := 0;
while i < n do
  i := i + 1;
  b ~~ {0,1};
  if b then binA++ else binB++ fi
end

  $\vphantom{\lstrnd}$
return (binA, binB) 
\end{lstlisting}
    \caption{Original programs $c_1, c_2$}
  \end{subfigure}
  \vrule
  \begin{subfigure}[t]{0.5\textwidth}
\begin{lstlisting}
i, binA, binB := 0;
while i < n $\land$ i < m do
  b ~~ {0,1};
  if b then binA++ else binB++ fi;
  i := i + 1;
end
while i < n do
  b ~~ {0,1};
  if b then binA++ else binB++ fi;
  i := i + 1;
end
return (binA, binB) 
\end{lstlisting}
    \caption{Intermediate program $c^*$}
  \end{subfigure}
  \caption{Coupling balls into bins}
  \label{fig:bib}
\end{figure}

The examples we have seen so far are all \emph{synchronous} couplings: they
relate the iterations of the while loop in lock-step. For some applications, we
may want to reason asynchronously, perhaps allowing one side to progress while
holding the other side fixed.  One example of an asynchronous coupling is
analyzing the \emph{balls into bins} process. We have two bins, and a set of $n$
balls. At each step, we throw a ball into a random bin, returning the count of
both bins when we have thrown all the balls. The code is on the left side in
\Cref{fig:bib}.

% \caption{Code for balls into bins}
% \label{fig:bib}
% \end{figure}

Now, we would like to consider what happens when we run two processes with
different numbers of balls. Intuitively, it is clear that if the first process
throws more balls than the second process, it should result in a higher load in
the bins; we aim to prove that the first process stochastically dominates the
second with the following coupling. Assume that the first process has more balls
($n_1 \geq n_2$). For the first $n_2$ balls, we have the two process do the same
thing---they choose the same bucket for their tosses. For the last $n_1 - n_2$
steps, the first process throws the rest of the balls. Evidently, this coupling
forces the bins in the first run to have higher load than the bins in the second
run.

To formalize this example, we again introduce a program $c^*$,
proving equivalence with $c_1$ and showing a coupling with $c_2$. The code for
$c^*$ is on the right side in \Cref{fig:bib}; we require the
dummy input $\lstt{m}$ to be equal to $\lstt{n}_2$.

Proving equivalence with program $c_1$ is direct, using the loop range splitting
transformation in EasyCrypt: $\WhileL{e}{c} \simeq \WhileL{e\wedge
  e'}{c};\WhileL{e}{c}$. Once this is done, we simply need to provide a coupling
between $c^*$ and $c_2$. By our choice of $\lstt{m}$, we can trivially couple
the first loop in $c^*$ to the (single) loop in $c_2$, ensuring that
$\post \triangleq \lstt{binA}^* \geq \lstt{binA}_2 \land \lstt{binB}^* \geq
\lstt{binB}_2$ after the first loop.

Then, we can apply the one-sided rules to couple the second loop in $c^*$
with a $\Skip$ statement in $c_2$. It is straightforward to show that $\post$ is
an invariant in rule \rname{WhileL}, from which we can conclude
$\Equiv{c^*}{c_2}{n_1 \geq n_2 \land m = n_2}{\lstt{binA}^* \geq \lstt{binA}_2
  \land \lstt{binB}^* \geq \lstt{binB}_2}$, and by equivalence of $c_1$ and
$c^*$ we have $\Equiv{c_1}{c_2}{n_1 \geq n_2}{\lstt{binA}_1 \geq \lstt{binA}_2
  \land \lstt{binB}_1 \geq \lstt{binB}_2}$, enough for stochastic domination by
\Cref{thm:sd}.

\section{Non-deterministic couplings: birth and death}

So far, we have seen \emph{deterministic} couplings, which reuse randomness from
the coupled processes in the coupling; this can be seen in the
\rname{Sample} rule, when we always choose a deterministic bijection. In this
section, we will see a more sophisticated coupling that injects new randomness. 
% \subsection{The ladder process}

% A simple example of a Markov process is a \emph{ladder process}. The process
% takes values in $\NN$. Starting at some initial position $x$, the process either
% increases by one (moves up the ladder), or resets to $0$ (falls off the ladder)
% every time step.  The probability of falling from location $i$ is $p_i$, with
% $p_i$ decreasing in $i$.

% \begin{figure}
%   \begin{center}
%     \includegraphics{ladder.pdf}
%   \end{center}
% \caption{Ladder processs}
% \end{figure}
% We can model this process with the following code:
% \jh{TODO}

% We would like to prove that if one process starts at a higher position than the
% other process, the first process will stochastically dominate the second
% process. Like before, we can prove this by finding an appropriate coupling.

% \jh{TODO explain coupling and put in invariant}

% \subsection{The birth and death chain}

For our example, we consider a classic Markov process. Roughly speaking, a
Markov process moves within a set of states each transition depending only on
the current state and a fresh random sample. The random walks we saw before are
classic examples of Markov processes.

A more complex Markov process is the \emph{birth and death chain}. The state
space is $\ZZ$, and the process starts at some integer $x$. At every time step,
if the process is at state $i$, the process has some probability $b_i$ of
increasing by one, and some probability $a_i$ of decreasing by one. Note that
$a_i$ and $b_i$ may add up to less than $1$: there can be some positive
probability $1 - a_i - b_i$ where the process stays fixed.

To model this process, we define
a sum type \lstt{Move} with three elements (\lstt{Left}, \lstt{Right} and
\lstt{Still}) which correspond to the possible moves a process can make. Then,
the chains are modeled by the code in the left of \Cref{fig:bd}, where
the distribution \lstt{bd(state)} is
the distribution of moves from \lstt{state}.

\begin{figure}
  \begin{subfigure}[t]{0.5\textwidth}
\begin{lstlisting}
H := []; state := start; i := 0;
while i < k do
  dir ~~ bd(state);
  $\vphantom{\lstrnd}$
  if dir = Left then
    state := state - 1;
  else if dir = Right then
    state := state + 1;
  fi
  H := state :: H;  
  i := i + 1;
end
return state
\end{lstlisting}
    \caption{Original programs $c_1, c_2$}
  \end{subfigure}
  \vrule
  \begin{subfigure}[t]{0.5\textwidth}
\begin{lstlisting}
H := []; state := start; i := 0;
while i < steps do
  d   ~~ dcouple;
  dir := proj [1|2] d;
  if dir = Left then
    state := state - 1;
  else if dir = Right then
    state := state + 1;
  fi
  H := state :: H;  
  i := i + 1;
end
return state
\end{lstlisting}
    \caption{Intermediate programs $c^*_1, c^*_2$}
  \end{subfigure}
  \caption{Coupling the birth and death chain}
  \label{fig:bd}
\end{figure}

Just like the biased coin and balls into bins processes, we want to prove
stochastic domination for two processes
started at states $\lstt{start}_1 \geq \lstt{start}_2$ via coupling. The
difficulty is that if the processes become adjacent and they both move,
the two processes may swap positions, losing stochastic domination.

The solution is to
use a special coupling when the two processes are on two adjacent states as in
\citet{mufa1994optimal}.
Unlike the previous examples, the coupling is not deterministic: the behavior of
one process is not fully determined by the randomness of the other.  
Our loop invariant is the usual one for stochastic domination:
$\lstt{state}_1 \geq \lstt{state}_2$.
To show that this invariant is preserved, we
perform a case analysis on whether $\lstt{state}_1 = \lstt{state}_2$,
$\lstt{state}_1 = \lstt{state}_2 + 1$ or $\lstt{state}_1 > \lstt{state}_2 + 1$.

We focus on the interesting middle case, when the states are adjacent. Here,
we perform a trick: we switch $c_1,
c_2$ for two equivalent intermediate programs $c^*_1, c^*_2$, and prove a
coupling on the two intermediate programs. The two intermediate programs each
sample from \lstt{dcouple}, a distribution on pairs of moves, and project out
the first or second component as \lstt{dir}; in other words, we explicitly
code $c^*_1, c^*_2$ as sampling from the two marginals of a common distribution
\lstt{dcouple}. By proving that the marginals are indeed distributed as
$\lstt{bd}(\lstt{state}_1)$ and $\lstt{bd}(\lstt{state}_2)$, we can prove
equivalences $c_1 \simeq c^*_1$ and $c_2 \simeq c^*_2$. The code is in the right
side of \Cref{fig:bd}, where
$\lstt{proj [1|2]}$ is the first and second projections in $c_1$ and $c_2$,
respectively.

All that remains is to
prove a coupling between $c^*_1$ and $c^*_2$ satisfying the loop invariant
$\lstt{state}_1 \geq \lstt{state}_2$.
With adjacent states, \lstt{dcouple} is given by the following function from pairs
of moves to probabilities:
\begin{lstlisting}
op distr-adjacent $a_i$ $a_{i+1}$ $b_i$ $b_{i+1}$ (x : Move * Move) =
  if x = (Right, Left ) then min($b_{i+1},a_{i}$)            else
  if x = (Still, Left ) then $(b_{i+1} - a_{i})^{+}$                 else
  if x = (Right, Still) then $(a_{i} - b_{i+1})^+$                 else
  if x = (Still, Right) then $a_{i+1}$                 else
  if x = (Left , Still) then $b_{i}$                 else
  if x = (Still, Still) then
    1 - min($b_{i+1}, a_{i}$) - $a_{i+1}$ - $b_{i}$ - $|b_{i+1}-a_{i}|$ else
  if x = (_    , _    ) then 0.
\end{lstlisting}
where $x^+$ denotes the \emph{positive part} of $x$: simply $x$ if $x \geq 0$,
and $0$ otherwise.
Note that the case \lstt{(Left, Right)} has probability $0$: this forbids the
first process from skipping past the second process.

Now the coupling is easy: we simply require both samples from \lstt{dcouple} to
be the same. Since $\lstt{state}_1 = \lstt{state}_2 + 1$ and the distribution
never returns $(\lstt{Left}, \lstt{Right})$, the loop invariant is trivially
preserved. This shows the desired coupling, and stochastic domination by
\Cref{thm:sd}.

\section{Conclusion and future work}

We have established the connection between relational verification of
probabilistic programs using pRHL, and probabilistic couplings. Furthermore, we
have used the connection by using pRHL to verify several well-known examples of
couplings from the literature on randomized algorithms.
More broadly, our work is a blend between the two main approaches
to relational verification: (i) reasoning about a single program combining the
two programs (e.g.
cross-products~\citep{ZaksP08}, self-composition~\citep{BartheDR04}, and product
programs~\citep{BartheCK11}); and (ii) using a program logic to reason directly
about two programs (e.g. relational Hoare logic~\citep{Benton04},
relational separation logic~\citep{Yang07}, and pRHL~\citep{BartheGZ09}).
We have only scratched the surface in verifying couplings; we see three natural
directions for future work.

\paragraph*{A more general verification framework.}
When we construct a coupling, the core data is encoded by the bijection $f$ for
the rule \rname{Sample}, which specifies how the two samples are to be coupled.
A careful look at the rule
  % \begin{center}
  %   \[
  %   \inferrule*[Left=\textrm{Sample}]{f\in T_1\stackrel{1-1}{\longrightarrow} T_2 \\
  %     \forall v\in T_1. ~d_1(v)=d_2(f~v)
  %   }{
  %     \Equiv{\Rand{x_1}{d_1}}{\Rand{x_2}{d_2}}
  %     {\forall v, \post [v/x_1\sidel,f(v)/x_2\sider]}
  %   {\post}}
% \]
  % \end{center}
reveals that the coupling is a \emph{deterministic} coupling, as defined by
\citet{Villani08}.
% \begin{definition}
%   A coupling $(X,Y)$ between two spaces $\mathcal{X},
%   \mathcal{Y}$ is said to be \emph{deterministic} if there exists a
%   function $T : \mathcal{X} \rightarrow \mathcal{Y}$ such
%   that $Y = T (X)$.
% \end{definition}
While such couplings are already quite powerful, there are
many examples of couplings that cannot be
verified using deterministic couplings. We have worked around
this difficulty by using program
transformation rules, but an alternative approach could be interesting:
allow
more general binary relations when relating samples, rather than just bijections.
This generalization could enable a more general class
of couplings and yield cleaner proofs.

Moreover, it would be interesting to extend \EasyCrypt with mechanisms
for handling the non-relational reasoning in couplings. To prove quantitative
bounds on total variation in the random walk example, we need to bound the time
it takes for a single random walk to reach a certain position. Proving such
bounds requires more complex, non-relational reasoning.
We are currently developing a program logic for this purpose, but it has
not yet been integrated into \EasyCrypt.

\paragraph*{Extending to shift and path coupling.}
The couplings realized in the random walks are instances of exact
couplings, where we reason about synchronized samples: we relate the first
samples, the second samples, etc. A more general
notion of coupling is \emph{shift-coupling},
where we are allowed to first shift one process by a random number of samples,
then couple.
The general theory of path
couplings provides similar-shaped inequalities as the ones in exact
coupling, allowing powerful mathematical-based reasoning inside the
logic with the \rname{Conseq} rule. These coupling notions are complex, and
it is not yet clear how they can be verified.

\paragraph*{Other examples.}
There are many other examples of couplings, in particular the proof of
the constructive Lovasz Local Lemma, a fundamental tool used in the
\emph{probabilistic method}, a powerful proof technique for showing
existence in combinatorics.

\subsubsection*{Acknowledgments.}
We thank Arthur Azevedo de Amorim and the anonymous reviewers for
their close reading and useful suggestions.  This work was partially
supported by a grant from the Simons Foundation (\#360368 to Justin
Hsu), NSF grant CNS-1065060, Madrid regional project S2009TIC-1465
PROMETIDOS, Spanish national projects TIN2009-14599 DESAFIOS 10
and TIN2012-39391-C04-01 Strongsoft, and a grant from the Cofund
Action AMAROUT II (\#291803).

\bibliographystyle{abbrvnat}
\bibliography{header,main}


%% \section{Graphical depictions of random processes}

%% We depict the two random walk processes in \Cref{fig:rw-pic,fig:torus-pic}.

%% \begin{figure}[t]
%% \begin{center}
%% \includegraphics{random_walk_1.pdf}
%% \end{center}
%% \caption{Unbiased random walk}
%% \label{fig:rw-pic}
%% \end{figure}

%% \begin{figure}[t]
%%   \begin{center}
%%     \includegraphics[height=3cm,width=10cm]{random_walk_2.pdf}
%%   \end{center}
%% \caption{Lazy random walk on a two dimensional torus}
%% \label{fig:torus-pic}
%% \end{figure}

\end{document}

%% file: prelude.tex
\usepackage[english]{babel}
\usepackage[T1]{fontenc}
\usepackage[utf8]{inputenc} 
\usepackage{upgreek}
\usepackage{paralist}
\usepackage{enumitem}
\usepackage{caption}
\usepackage{subcaption}
\usepackage[numbers,longnamesfirst]{natbib}
\usepackage{xspace}
\usepackage{amsmath,amsfonts,amssymb}
\usepackage{xcolor}
\usepackage[ruled,vlined]{algorithm2e}
\usepackage{nicefrac}
\usepackage{bbm}
\usepackage{graphicx}
\newcommand{\EasyCrypt}{\textsf{EasyCrypt}\xspace}

\usepackage{mathpartir}
\usepackage{yfonts}

\usepackage{todonotes}

\usepackage{xargs}
\newcommand{\rname}[1]{[{\sc #1}]}
\newcommand{\Dist}{\ensuremath{\mathbf{Distr}}}

\newcommand{\bernD}{\mathbf{Bern}}

\newcommand{\ZZ}{\mathbb{Z}}

\newcommand{\Mem}{\mathsf{Mem}}

\newcommand\q{[\![}
\newcommand\p{]\!]}
\newcommand\Sem[1]{\q #1 \p}

\newcommand{\Skip}{\mathsf{skip}}
\newcommand{\Seq}[2]{#1;\ #2}
\newcommand{\WhileL}[2]{\mathsf{while}\ #1\ \mathsf{do}\ #2}
\newcommand{\Cond}[3]{\mathsf{if}\ #1\ \mathsf{then}\ #2\ \mathsf{else}\ #3}

\newcommand{\Ass}[2]{#1 \leftarrow #2}

\newcommand{\Rand}[2]{#1 \stackrel{\raisebox{-.25ex}[.25ex]{\tiny $\mathdollar$}}{\raisebox{-.2ex}[.2ex]{$\leftarrow$}} #2}

\renewcommand{\Pr}[2]{\mathrm{Pr}_{#1}{\left[ #2 \right]}}

\newcommand{\pre}{\Psi}
\newcommand{\post}{\Phi}

\newcommand{\Equiv}[4]{%
  \vDash #1 \sim {#2}: {#3} \Rightarrow {#4} }

\usepackage{listings}

\usepackage[T1]{fontenc}
\usepackage{microtype}

\usepackage[scaled]{beramono}
\newcommand\Small{\fontsize{8.2pt}{8.4pt}\selectfont}
\newcommand*\LSTfont{\Small\ttfamily\SetTracking{encoding=*}{-60}\lsstyle}
\def\lstrnd{\stackrel{\raisebox{-.15ex}{\ensuremath{\scriptscriptstyle\$}}}{\raisebox{-.2ex}{\ensuremath{\leftarrow}}}}

\newcommand{\lstt}[1]{\mbox{\LSTfont #1}}

\usepackage{xcolor}
\definecolor{DarkGreen}{rgb}{0.1,0.5,0.1}
\definecolor{DarkRed}{rgb}{0.5,0.1,0.1}
\definecolor{DarkBlue}{rgb}{0.1,0.1,0.5}
\usepackage{hyperref}
\hypersetup{
    unicode=false,          % non-Latin characters in Acrobat's bookmarks
    pdftoolbar=true,        % show Acrobat toolbar?
    pdfmenubar=true,        % show Acrobat menu?
    pdffitwindow=false,      % page fit to window when opened
    pdftitle={},    % title
    pdfauthor={}
    pdfsubject={},   % subject of the document
    pdfnewwindow=true,      % links in new window
    pdfkeywords={keywords}, % list of keywords
    colorlinks=true,       % false: boxed links; true: colored links
    linkcolor=DarkRed,          % color of internal links
    citecolor=DarkGreen,        % color of links to bibliography
    filecolor=DarkRed,      % color of file links
    urlcolor=DarkBlue,          % color of external links
}

\usepackage{cleveref}